\documentclass[12pt]{article}
\usepackage[margin=0.95 in]{geometry}
\usepackage{amsmath}
\usepackage{amssymb,amsfonts}
\usepackage[all]{xy}
\usepackage{graphicx}
\usepackage[utf8x]{inputenc}
\usepackage{amsmath}
\usepackage{amssymb}
\usepackage{float}
\usepackage{array}
\usepackage{tikz}
\usepackage{mathtools}
\usepackage{mathrsfs}
\usepackage{hyperref}
\usepackage{cite}
\numberwithin{equation}{section}
\numberwithin{equation}{section}
\numberwithin{table}{section}
\begin{document}

\thispagestyle{empty}

\vspace*{3cm}
{}
\noindent
{\LARGE \bf  A Duality in Two-Dimensional Gravity}
\vskip .4cm
\noindent
\linethickness{.06cm}
\line(10,0){467}
\vskip 1.1cm
\noindent
\noindent
{\large \bf Sujay K. Ashok$^a$ and Jan Troost$^b$}
\vskip 0.5cm
{\em 
\noindent
$^a$Institute of Mathematical Sciences \\
Homi Bhabha National Institute (HBNI)\\
IV Cross Road, C.~I.~T.~Campus, \\
  Taramani, Chennai, 600113  Tamil Nadu, India}
\vskip 0.5cm
{\em
\noindent
$^b$Laboratoire de Physique Th\'eorique
de l'\'Ecole Normale Sup\'erieure \\
\hskip -.05cm
 CNRS,
 PSL Research University, Sorbonne Universit\'e \\
 Paris, France
}
\vskip 1.2cm

\vskip0cm

\noindent {\sc Abstract: }  We demonstrate an equivalence
between two integrable flows defined in a polynomial ring quotiented by an ideal generated by a polynomial. This duality of integrable systems allows us to systematically exploit
the Korteweg-de Vries   hierarchy and its tau-function to propose amplitudes for non-compact  topological gravity on Riemann surfaces of arbitrary genus. We thus quantize topological gravity coupled to non-compact topological matter and demonstrate that this phase of topological gravity at $N=2$ matter central charge larger than three is equivalent to the phase with matter of central charge smaller than three. 

\vskip 1cm

\pagebreak

\newpage
\setcounter{tocdepth}{2}
\tableofcontents

\section{Introduction}
Field theories with $N=2$ supersymmetry in two dimensions give rise to  topological quantum field theories after twisting \cite{Witten:1988ze}. When the starting point
is a non-compact conformal field theory, the  correlation functions of the resulting topological quantum field theories were  recently computed \cite{Li:2018rcl}. Subsequently, these theories were  coupled to  topological gravity \cite{Li:2018quf,Witten:1989ig} and the gravitational theory was solved on the sphere.\footnote{For reviews of theories of gravity in two dimensions,
see \cite{Ginsparg:1993is,DiFrancesco:1993cyw,Dijkgraaf:1990qw,Dijkgraaf:1991qh}.} 

One motivation for studying topological gravity coupled to non-compact matter is to test the gravitational consequences
of going beyond the central charge bound $c=3$ for $N=2$ minimal matter in two dimensions. In \cite{Li:2018quf}, it was noted that
coupling topological gravity to twisted matter with central charge $c>3$ gives rise to critical behavior reminiscent of $N=2$ matter with central charge $c<3$ coupled to  gravity. We will gain more insight into this  similarity in the present paper.
Another motivation for the study of these theories is the wish to compute topological string amplitudes on asymptotically linear dilaton spaces which are
generalizations of non-compact Calabi-Yau manifolds \cite{Ooguri:1995wj,Hori:2001ax,Hori:2002cd,Giveon:1999px,Eguchi:2000tc,Ashok:2007ui}.

In this paper, we exhibit the close relation between non-compact topological quantum field theories \cite{Li:2018rcl,Li:2018quf} and the deformations of topologically twisted compact $N=2$ minimal models \cite{Dijkgraaf:1990qw}. By detailing the link, we also gain control over the non-compact topological quantum field theories coupled to topological gravity on Riemann surfaces of higher genus. We thus extend their solution at genus zero \cite{Li:2018quf} to arbitrary genera.

The underlying idea of the equivalence is simple. An $N=2$ minimal model is the infrared fixed point of a Landau-Ginzburg theory in two dimensions with $N=(2,2)$ supersymmetry and a single chiral superfield, subject to superpotential interactions. The minimal model with central charge $c=3-6/k_c$ corresponds to a superpotential monomial
$W_c=X^{k_c}$ where $X$ is a $N=(2,2)$ chiral superfield. On the other hand, a $N=2$ non-rational conformal field theory at central charge $c=3+6/k$, with $k$ a positive integer can be modelled with a generalized Landau-Ginzburg theory with superpotential $Y^{-k}$, where $Y$ is again a chiral superfield \cite{Ooguri:1995wj,Hori:2001ax,Li:2018rcl}. The topological quantum field theories we study are the theories that arise upon twisting and deforming the infrared fixed points.  Formally, the change of variables $X=Y^{-1}$ maps the superpotential of the compact model to the superpotential of the non-compact model (upon identifying the levels $k_c=k$).
In this paper, we analyze the extent to which the change of variables proves an equivalence between the compact and non-compact topological quantum field theories, and their coupling to gravity.

The correlation functions of both the compact and non-compact models are governed by the Korteweg-de Vries (KdV) or reduced Kadomtsev-Petviashvili (KP) integrable hierarchy.\footnote{See e.g. \cite{Dickey:1991xa} for an introduction to these hierarchies.}  Our duality comes down to chasing the change of variables $X=Y^{-1}$ through the
equations determining the classical and quantum integrable hierarchy. Thus, the calculational proofs are elementary. 
 We then  move to exploit 
this duality to solve  non-compact topological gravity in two dimensions on Riemann surfaces of any genus. The duality  provides a technically  transparent though
conceptually challenging answer to a  hard question in two-dimensional quantum gravity, which pertains to the backreaction of gravity in response to a large amount of matter.

Similar (though not identical) ideas have been mentioned in the integrable hierarchy literature. Firstly, there is an equivalence relation that was mentioned for the classical 
rational KP integrable hierarchy in \cite{Aoyama:1994nb}. It was applied  to non-polynomial examples. Secondly, a similar device was employed to argue that matrix models with negative
power monomial potential are governed by a reduced KP hierarchy \cite{Mironov:1994mv}. Thirdly, we observe that the inversion of variables comes down to an
analytic continuation of the exponent of the superpotential from $k_c$ to $-k$, where both $k_c$ and $k$ are positive integers.\footnote{Note that we  have equality of the levels under the duality map. We will come back to the relation between these approaches in section \ref{analytic}.} This analytic continuation was 
proposed as a  method for obtaining results about models of topological gravity coupled to a topological non-compact coset conformal field theory \cite{Witten:1991mk}, or matrix models with a negative power monomial
potential \cite{Brezin:2012uc}. 

Our plan is to kick off the paper by proving the equivalence between two dispersionless (i.e. classical, tree level, spherical) integrable hierarchies in section \ref{classicalequivalence}. 
In section \ref{strictlimit} we detail the relation between the non-compact solution obtained by duality and the solution to the model obtained by
twisting the physical spectrum of $N=2$ Liouville theory studied in
\cite{Li:2018rcl,Li:2018quf}.
We then solve the non-compact topological quantum field theory coupled to gravity using a dispersionful (or quantum) 
KdV hierarchy in section \ref{quantumsolution}.
In section \ref{analyticcontinuation},
we discuss the extent to which our solution relates to the  approach of determining the 
correlators of a non-compact model through analytic continuation in the central
charge (or level) of the conformal field theory \cite{Witten:1991mk,Brezin:2012uc}. We conclude in section \ref{conclusions} and comment on the conceptual implications of the duality for two-dimensional gravity. In appendices \ref{illustrations} and \ref{modelsandscaling}
 we provide illustrations that may help to reveal aspects of our paper as either subtle, or simple.

\section{The  Duality}
\label{duality}
\label{classicalequivalence}
In this section, we briefly review the rational Kadomtsev-Petviashvili (KP) hierarchy reduced with respect to the derivative of a (super)potential.
We follow the pedagogical reference \cite{Aoyama:1994nb} and refer to \cite{Dickey:1991xa} for background. Importantly, the hierarchy can be formulated democratically with respect to the times of the integrable evolutions. Moreover, we carefully choose our setting  sufficiently broadly to allow for all manipulations that we will need. Then, we prove an equivalence between a model with polynomial potential and a model which is polynomial in the inverse variable. These are the integrable hierarchies corresponding to a compact topological quantum field theory and a non-compact topological quantum field theory respectively (see  e.g. \cite{Dijkgraaf:1990qw} and \cite{Li:2018rcl} and references therein for the relation to deformations of supersymmetric
conformal field theories and the operation of twisting).  We prove the classical equivalence of these models in this section, and discuss  the quantum, dispersionful hierarchy in section \ref{quantum}.

\subsection{The Rational KP Hierarchy in a Nutshell}
We very briefly review the rich rational KP hierarchy. We  must refer to \cite{Aoyama:1994nb} for  more background information and a laundry list of intermediate results. The basic data of the hierarchy is a potential $W$ which is a polynomial in the ring $\mathbb{C}[X,X^{-1}]$.\footnote{In the context of $N=(2,2)$ supersymmetric
Landau-Ginzburg models with chiral superfield $X$, the potential  is identified with the superpotential  $W(X)$.}
It has a minimal and maximal degree.\footnote{We ignore the constant term in determining these degrees since it is the derivative of the (super)potential that will be crucial.} If  the maximal degree $k_{max}$ is positive and the minimal degree $-k_{min}$ is negative, we define two formal power series, one  in $X^{-1}$, and one in $X$, through the formulas
\begin{eqnarray}
L_{max} &=& \left(\frac{W}{c_{max}} \right)^{\frac{1}{k_{max}}} = X \, \, + O(1)+O(X^{-1})+ \dots
\nonumber \\
L_{min} &=& (k_{min} W)^{\frac{1}{k_{min}}} = \left(c_{min}k_{min} \right)^{\frac{1}{k_{min}}} X^{-1} + O(1)+O(X)+\dots \, .
\end{eqnarray}
We normalized the first formal power series (by dividing by the leading coefficient $c_{max}$ of the polynomial $W$) such that the first term on the right hand side has coefficient one, and we will often pick
$c_{min}=1/k_{min}$. Next, we define Hamiltonians $Q^i$ through the formulas:
\begin{eqnarray}
Q^{i \ge 0} &=& \frac{1}{i+1}\left[ L_{max}^{i+1}\right]_{\ge 0}
\nonumber \\
Q^{i \le -2} &=&  \frac{1}{i+1} \left[L_{min}^{-i-1}\right ]_{\le -1} \, ,
\end{eqnarray}
where the indices on the square brackets indicate which orders in the formal power series we keep.\footnote{There exists an important extension to include
a Hamiltonian $Q^{-1}$, but we barely need it in this paper.} We introduce an infinite set of times $t_{i \in \mathbb{Z} \setminus \{ -1 \} }$
 and the reduced integrable hierarchy is then defined  by the evolution equations:
\begin{equation}
\{ Q^j, W \}_{t_i} = \{ Q^i, W \}_{t_j} \label{evolution}
\end{equation}
which is shorthand for 
\begin{equation}
\partial_X Q^j \partial_{t_i} W - \partial_X W \partial_{t_i} Q^j
= \partial_X Q^i \partial_{t_j} W - \partial_X W \partial_{t_j} Q^i \, .
\label{evolutionexplicit}
\end{equation}
Because the connection $Q$ is flat, the time evolutions in the  ring are mutually compatible \cite{Aoyama:1994nb}.
We chose a formulation of the hierarchy which is democratic with respect to all time variables. Given the integrable hierarchy, one can  define a set of operators, topological quantum field theory correlators, as well as classical topological gravitational correlators that satisfy all the  axioms of such theories (such as associativity of the operator product and a topological recursion relation). Moreover, generating functions for these correlators can be constructed. See e.g. \cite{Aoyama:1994nb} for the large set of standard, relevant and explicit formulas.
We assume these topological quantum field theories and topological theories of gravity to be known in the following.

\subsection{The Compact and Non-Compact Flows}
After the brief recap of the general framework of the rational Kadomtsev-Petviashvili hierarchy, we simplify matters considerably in this section. We concentrate on proving and analyzing in detail a duality between two reduced KP hierarchies. The first is
the integrable KP hierarchy reduced over the derivative of  a polynomial (super)potential in a variable $X$
and the second is a theory reduced over the derivative of a polynomial potential in a variable $Y^{-1}$. The existence of such a duality for the strictly rational case is mentioned in  \cite{Aoyama:1994nb}. The necessity of introducing the rational framework despite the polynomial nature of our potentials lies
in the fact that we wish to be able to divide by polynomials in the following.

Our two families of models can be described explicitly as follows.
The compact model parameterised by the variable $X$ has a polynomial potential
\begin{equation}
W_c(X)= \frac{X^{k_c}}{k_c} + v_{k_c-2} X^{k_c-2} + 
\dots + v_1 X + v_0 \, ,
\label{compactsuperpotential}
\end{equation}
while the non-compact model with variable $Y$ has a polynomial potential in $Y^{-1}$
\begin{equation}
W=v_0 + v_{-1} Y^{-1} + \frac{v_{-2}}{2}  Y^{-2} + \dots + \frac{v_{-k+2} }{k-2}Y^{-k+2} + \frac{1}{k} Y^{-k} \, .
\label{noncompactsuperpotential}
\end{equation}
The normalization of the coefficients is chosen to agree with \cite{Aoyama:1994nb}.\footnote{The notation $v_0$ is used twice, but since these coefficients will soon be mapped into each other, we hope this convenient abuse of notation will not confuse the reader.}
In the following it is important to assume that the coefficients $v_1$ as well as $v_{-1}$ are non-zero since these determine the
minimal and maximal degree of the derivative  of the potential respectively, and consequently the dimension of the quotient ring.
The leading coefficient of the potentials is chosen to be fixed and non-zero. The subleading coefficient of the potentials is chosen to be zero through a shift of the variables $X$, respectively $Y^{-1}$.

 For both the compact and the non-compact models, we can 
define generators $Q^{i \in \mathbb{Z}}$ of KP integrable flows parameterized by times $t_{i \in \mathbb{Z}}$.
For the compact models only the Hamiltonians with $i \ge -1$ are non-trivial, while for the non-compact models those with
$i \le -1$ are non-trivial. For simplicity, we again exclude the time $t_{-1}$ from our considerations in both models, though it can 
be reinstated if so desired. 

We label quantities referring to the compact model with an extra index '$c$' while those quantities without extra index relate to the non-compact model. Thus, for the compact model we have times $t^c_{i \ge 0}$ and for the non-compact model
times $t_{i \le -2}$. To restate the time evolutions, we introduce the roots of the potential for the compact and the non-compact model
as formal power series at large $X$ and small $Y$ respectively:
\begin{eqnarray}
L_c &=& (k_c W_c)^{\frac{1}{k_c}} = X + O\left(\frac{1}{X}\right)
\label{compactroot}
 \\
L &=& (k W)^{\frac{1}{k}} = Y^{-1} + O(Y) \, .
\label{noncompactroot}
\end{eqnarray}
The Hamiltonians are 
\begin{eqnarray}
Q_c^i &=& \frac{1}{i+1} \left[L_c^{i+1}\right]_{\ge 0}
\label{compactHamiltonians} \\
Q^i &=&  \frac{1}{i+1}\left[ L^{-i-1} \right]_{\le -1} \, .
\label{noncompactHamiltonians}
\end{eqnarray}
The reduced integrable hierarchies are then defined  by the evolution equations
(\ref{evolution}), valid for both the compact and the non-compact models.
In the compact model, the evolution equations take a standard form if we pick $t_j^c=t_0^c$. They then read:
\begin{equation}
\partial_{t_i^c} W_c = \{ Q_c^i,W_c \} = \partial_X Q_c^i \partial_{t_0^c} W_c - \partial_{t_0^c} Q_c^i \partial_X W_c \, .
\label{compactevolution}
\end{equation}
In the non-compact model,
if we pick the first possible time, $t_j=t_{-2}$ as our reference time, than the equations defining the hierarchy become
\begin{equation}
\partial_{t_i} W = Y^2 ( \partial_Y Q^i \partial_{t_{-2}} W - \partial_Y W \partial_{t_{-2}} Q^i) \, ,  \label{noncompactevolution}
\end{equation}
because we have the equality $\partial_Y Q^{-2}=Y^{-2}$. In other words, we find a different symplectic
structure.

\subsection{The Equivalence of  Flows}

The first part of our equivalence map is to demonstrate that the classical flows of the compact and non-compact integrable hierarchies are isomorphic.
We relate the flows of these two reduced KP hierarchies through the change of variables $X=Y^{-1}$.
If we  equate the levels
$k_c=k$ in both models as well as the potential coefficients $v_{a}=v_{-a}/a$ for $a \in \{ 1, 2, \dots , k_c-2 \}$,
then after the change of variables $X=Y^{-1}$, the compact superpotential $W_c$
(\ref{compactsuperpotential}) and the non-compact superpotential $W$ (\ref{noncompactsuperpotential}) match. Thus,
the series expansions $L_c $ (\ref{compactroot}) and $L$ (\ref{noncompactroot}) are equal. We moreover map the indices $i_c+1 \leftrightarrow -i-1$ which implies $ i_c \leftrightarrow -i-2$,
to match the compact Hamiltonians $Q_c^{i \ge -1}$ (\ref{compactHamiltonians}) with the negative of the non-compact Hamiltonians $Q^{ i \le -1}$ (\ref{noncompactHamiltonians}). Most importantly, we note that the times $t_i^c$ are mapped to the times $t_{-i-2}$, and that under the change of variables $X=Y^{-1}$, the flow evolution equations
(\ref{compactevolution}) and (\ref{noncompactevolution}) 
are mapped to each other, because $\partial_X = - Y^2 \partial_Y$. The overall sign works out
as well because of the minus sign in the comparison of Hamiltonians. The symplectic structures map
into each other. We conclude that the integrable flows agree under the duality map. 

For the reader's convenience, we summarize the substitution rules:
\begin{eqnarray}
X  &\longleftrightarrow&  Y^{-1}   \quad \qquad \, \, \,
t_i^c  \longleftrightarrow  t_{-i-2}
\nonumber \\
i^c  &\longleftrightarrow&  - i -2
 \qquad 
Q^i_c  \longleftrightarrow  -Q^{-i-2}
\nonumber \\
L_c  &\longleftrightarrow&  L \, .   \label{equivalencemap}
\end{eqnarray}

\subsection{The Operator Rings}
We matched the integrable hierarchies. We now work out further details of how the operator rings and other data match between the topological quantum field theories as well as the theories of topological gravity. For each model, an infinite set of operators is defined as derivatives of the
Hamiltonians:
\begin{eqnarray}
\phi^i_c &=& \partial_X Q^i_c \, ,
\nonumber \\
\phi^i &=& \partial_Y Q^i \, .
\end{eqnarray}
These operators live in the rings $\mathbb{C}[X,X^{-1}]$ and $\mathbb{C}[Y,Y^{-1}]$ respectively. Due to the condition that the coefficients $v_{\pm 1}$ are non-zero,  the quotient rings where we divide by the ideal generated by the derivative of the superpotential
have bases $\phi^{\alpha}$ where $\alpha \in \Delta_c = \{ 0, 1, \dots, k_c-2 \}$ for the compact model and
$\alpha \in \Delta = \{ -2, -3, \dots, -k \}$ for the non-compact model. These quotient rings will be more manifestly isomorphic under duality after making a  change of basis of the type discussed in \cite{Aoyama:1994nb}. In the non-compact model, we pick a reference basis element $\phi^{\alpha_0=-2}$, and divide all operators in the basis by this reference element. Firstly, we recall that
\begin{equation}
Q^{-2} = - [ L ]_{\le -1} = - Y^{-1} \, ,
\end{equation}
and therefore the corresponding operator $\phi^{-2}$ is
\begin{equation}
\phi^{-2} = Y^{-2} \, .
\end{equation}
We pick the operator basis $\tilde{\phi}^{\alpha}$ in the non-compact model given by
\begin{equation}
\tilde{\phi}^\alpha = \frac{\phi^\alpha}{\phi^{\alpha_0}} = Y^{2} \phi^\alpha \, .
 \end{equation}
In fact, we can more generally define new operators\footnote{Here, we go beyond our background reference \cite{Aoyama:1994nb}.}
\begin{eqnarray}
\tilde{\phi}^i &=& \frac{\phi^i}{\phi^{-2}} = Y^2 \partial_Y Q^i
\end{eqnarray}
which under the equivalence map (\ref{equivalencemap}) map onto the operators $\phi^{-i-2}_c$ of the compact model.


\subsection{The Topological Quantum Field Theories}
The topological quantum field theories associated to the equivalent integrable systems are necessarily isomorphic. There are some subtleties in the details of the matching that we want to discuss.
The references \cite{Aoyama:1994nb,Li:2018quf} define universal coordinates for the non-compact models. They are useful, and allow for an explicit solution of the non-compact quantum field theory correlators \cite{Li:2018rcl,Li:2018quf}. They are, however, different from the universal coordinates one finds under the equivalence map from the universal compact coordinates. In this paper, we  work in the latter coordinates, since we are focused on exploiting the equivalence map. To make the relation between the results in this paper and those in \cite{Aoyama:1994nb,Li:2018quf} more manifest, we 
 record the explicit map between the universal coordinates we use in this
subsection to those used in the original description of the non-compact model \cite{Aoyama:1994nb,Li:2018quf}
in appendix \ref{universalcoordinatesmap}. Below, we  define the new non-compact universal coordinates that are natural from the perspective of the duality.

The compact universal coordinates $u_c$ are described by inverting the series $L_c(X)$:
\begin{equation}
X = L_c - \frac{u^0_c}{L_c} - \dots - \frac{u_c^i}{L^{i+1}_c} - \dots 
\end{equation}
They coincide with the $i0$ component of the Gelfand-Dickey potentials $G^{ij}_c$ \cite{Aoyama:1994nb}
\begin{equation}
G^{ij}_c = \frac{1}{i+1} \text{Res}_{X=\infty} (L^{i+1}_c \phi^j_c) \, ,
\end{equation}
namely
\begin{equation}
u^i_c = G^{i0}_c \, .
\end{equation}
For the non-compact theory on the other hand, we define the potentials \cite{Aoyama:1994nb}
\begin{equation}
G^{ij} = -\frac{1}{i+1} \text{Res}_{Y=0} \left( L^{-i-1} \phi^j \right)
\end{equation}
and the new universal coordinates $\tilde{u}^i$ adapted to the choice $\alpha_0=-2$:
\begin{equation}
\tilde{u}^i = G^{i,-2} \, .
\end{equation}
We can prove that the universal coordinates $u_c$ and $\tilde{u}$ do match under the equivalence map, using the following property of residue formulas.
The relation between the residue of a function or formal power series at infinity and the residue at zero is (in conventions in which both are defined as the coefficient of the term with power minus one):
\begin{equation}
\text{Res}_{X=\infty} f(X) = \text{Res}_{Y=0} \left(Y^{-2} f(Y^{-1})\right) \, . \label{residueformula}
\end{equation}
Using this property, the universal coordinates agree (since the roots of the superpotentials, the indices and the operators are appropriately mapped).

It is then straightforward to follow all the quantities that determine the 
topological quantum field theory through the equivalence map. 
Indeed, under the substitution rules, we have more generally:
\begin{equation}
G^{ij}_c(u_c) = G^{-i-2,-j-2}(\tilde{u}) \, .
\end{equation}
Since these are second derivatives of the generating functions \cite{Aoyama:1994nb},
we can integrate the equality up and
pick generating functions of correlation functions of matter correlation functions $F_m$ that are equal 
\begin{equation}
F^{(0)}_{c,m}(u^c) = F_m^{(-2)}(\tilde{u}) \, .
\end{equation}
The upper index refers to the picture in which we evaluate the correlation functions, which is the
zero picture for the compact model, and the minus two picture for the non-compact model (since for the
latter we chose the reference time $t_{-2}$).
This proves the equality of universal $N$-point functions, defined as derivatives with respect to $u_c$
and $\tilde{u}$ coordinates respectively. In other words, we have
\begin{align}
\langle \phi^{\alpha_1}_c \dots \phi^{\alpha_N}_c \rangle_0 &= \partial_{u^c_{\alpha_1}} \dots 
\partial_{u^c_{\alpha_N}} F^{(0)}_{c,m}(u^c)\cr 
&= \partial_{\tilde{u}_{\alpha_1}} \dots
\partial_{\tilde{u}_{\alpha_N}} F_m^{(-2)}(\tilde{u}) = \langle \tilde{\phi}^{\alpha_1} \dots \tilde{\phi}^{\alpha_N}
\rangle_{\alpha_0=-2} \, .
\end{align}
To illustrate the equivalence hands-on, we note that the metrics
$\eta_{\alpha \beta}^c= \delta_{\alpha+\beta,k_c-2}$ and $\eta_{\alpha \beta} = \delta_{\alpha+\beta,-k-2}$
match since if we have primaries $\alpha,\beta$ and $\alpha+\beta=k_c-2$ before the transformation,
then $-\alpha-2-\beta-2=-k-2$ after the transformation. The structure constants also match.
The first reason for this is that the operators $\phi^\alpha_c$ and $\tilde{\phi}^\alpha$ are bases of the respective
rings that agree on the nose under the equivalence map (as shown in the previous subsection). Thus, the corresponding structure constants will automatically coincide. 
This is sufficient to prove that the associated topological quantum field theories match on all Riemann surfaces, in agreement
with the equality of generating functions.

Let us  provide even more details on the correspondence of the  formulas. 
We can show how a scalar product on the space of operators agrees between the compact model with respect to time $t_0$ and the non-compact
model with respect to time $t_{-2}$.
For the compact model, we have the scalar product \cite{Aoyama:1994nb}
\begin{equation}
(\phi_c, \chi_{c})_{0} \equiv \text{Res}_{X \in \text{Ker} W^c_X} \left(\frac{\phi_c \chi_{c}}{W^c_X} \right) \, ,
\end{equation}
where we sum over the zeroes of the derivative of the superpotential $W^c$.
For the non-compact model, with the reference choice  $\alpha_0=-2$, we define the scalar product \cite{Aoyama:1994nb}:
\begin{equation}
(\tilde{\phi},\tilde{\psi})_{\alpha_0} = \text{Res}_{Y \in \text{Ker} W_Y} \left( 
  \frac{\tilde{\phi}\, \tilde{\psi}\, (\phi^{\alpha_0})^2}{W_Y} \right) \, .
  \label{ncscalar}
\end{equation}
Firstly, this  manifestly implies that $(\phi,\psi)_0=(\tilde{\phi},\tilde{\psi})_{\alpha_0}$, by the definition of the operators $\tilde{\phi}$.
 Secondly, we can sum over the zeroes of the derivative of the superpotential, which for the compact model lie near zero by performing a large contour integral. We can slip this contour over the sphere in the compact model, and evaluate it at $X=\infty$. For the non-compact model, we reason similarly and can evaluate the residue formula at $Y=0$. Using this contour deformation, we find under the equivalence map and using property (\ref{residueformula}) that
\begin{equation}
(\tilde{\phi},\tilde{\chi})_{-2}= 
(\phi_c,\chi_c)_0 \,  \, .
\end{equation}
The structure constants similarly match  since (see \cite{Aoyama:1994nb} for the proof of the residue formula for the three-point correlator):
\begin{eqnarray}
{{\tilde{c}}^{\alpha \beta}}_{\, \, \, \, \, \gamma}
&=& \langle \tilde{\phi}^\alpha \tilde{\phi}^\beta \tilde{\phi}_\gamma  \rangle_{\alpha_0} = 
\text{Res}_{Y \in Ker W_Y}  \left( \frac{\tilde{\phi}^\alpha \tilde{\phi}^\beta \tilde{\phi}_\gamma (\phi^{\alpha_0})^2}{  W_Y}  \right)
\nonumber \\
&=& \text{Res}_{X \in Ker W_X^c}  \left(\frac{{\phi}^\alpha_c{\phi}^\beta_c {\phi}_\gamma^c}{ W_X^c}  \right) =
\langle \phi^\alpha_c \phi^\beta_c \phi_\gamma \rangle_0=
 {{c_c}^{\alpha \beta}}_\gamma \, . \label{ncthree}
\end{eqnarray}
Let us remark that we can interpret the $\phi^{\alpha_0}$ insertions in formulas (\ref{ncscalar}) and
(\ref{ncthree}) as taking us from a zero picture vacuum state to a minus two picture vacuum state where the
two are related by $| \mbox{vac} \rangle_{(-2)} = \phi^{-2} | \mbox{vac} \rangle_{(0)}$.
Finally, we backtrack to the time evolution of the coefficients of the superpotential with the integrable
flows parameterized by the times. The reduced KP hierarchy has a 
topological quantum field theory solution $\tilde{u}^\alpha = \eta^{\alpha \beta} t_\beta$ when we restrict the range
of times to the basis set $\Delta$ (for either the compact or the non-compact theory) \cite{Aoyama:1994nb}.
These topological quantum field theory solutions also map from the compact to the non-compact problem under the substitution
$u^\alpha_c \leftrightarrow \tilde{u}^{-\alpha-2}$.

In summary, we find that a compact topological quantum field theory captured by a polynomial superpotential is equivalent to a non-compact topological quantum field theory defined by a superpotential of an inverse variable.
The relation between the theories is rather involved in terms of the 
standard universal coordinates (see appendix \ref{universalcoordinatesmap}), and becomes straightforward in the universal coordinates suggested by the equivalence map.

\subsection{The Classical Gravitational Equivalence}

We have proven a classical equivalence  of a compact and a non-compact topological quantum field theory. In this subsection, we provide only a few of the details of the equivalence of the topological quantum field theories coupled to topological gravity, at the classical level, i.e. on the sphere. The idea of the map is again simple. We extend the agreement of the flows labelled by $\alpha$ exploited in the topological quantum field theory equivalence to include all the times of the integrable hierarchy. We need to go slightly beyond the 
discussion provided in \cite{Aoyama:1994nb} on this occasion.

There is an equivalence map for descendants fields. We define \cite{Aoyama:1994nb}
\begin{eqnarray}
\sigma_N (\phi^\alpha_c) &=& \left( (|\alpha+1|)(|\alpha+1|+k_c) \dots (|\alpha+1| + (N-1) k_c) \right)^{-1} \phi^{N k_c+\alpha}_c
\nonumber \\
\sigma_N (\tilde{\phi}^\alpha) &=& \left( (|\alpha+1|)(|\alpha+1|+k) \dots (|\alpha+1|+(N-1)k) \right)^{-1}  \tilde{\phi}^{-N k + \alpha} \, .
\end{eqnarray}
Under the equivalence map, these operators and their normalisations match.
Since the $\tilde{\phi}^\alpha$ operators are an alternative basis for the quotient ring, the descendant decomposition theorem  (which says that any descendant can be decomposed into primaries) as well as the topological recursion relation for the operators $\sigma_N$ (which says that descendant three-point functions can be recursively computed in terms of primary three-point functions -- see \cite{Witten:1989ig,Dijkgraaf:1990qw,Dijkgraaf:1991qh} for background --) are valid also in the $\alpha_0=-2$ reference frame.
Moreover, both theorems are mapped to their compact counterparts under the equivalence map. Thus, the equivalence map extends to the topological quantum field theories on the sphere coupled to gravity.
The compact generating function of gravitational correlators is also mapped to its non-compact counterpart.

\subsubsection*{Summary}

The duality map is now manifest. We used the change of variables $X \leftrightarrow Y^{-1}$ to prove the equivalence of the potentials $Q$ and $Q_c$, of the fields $\phi_c$ and $\tilde{\phi}$, of the times $t_c^i$ and $t^i$,
of the universal coordinates $u_c$ and $\tilde{u}$, and finally of the Gelfand-Dickey potentials $G$ and generating functions $F$.
Thus the full classical equivalence is understood.  
The statement that we obtain is the following. If we consider an extended non-compact model with a superpotential of the form (\ref{noncompactsuperpotential}),
then a choice of reference time $t_{-2}$ 
provides a model   equivalent to the compact model (\ref{compactsuperpotential}) in the more standard reference time $t_0$.
In the next section, we study  how the trivialization of this  non-compact model is related to the solution of the twisted $N=2$ Liouville theory obtained in \cite{Li:2018rcl,Li:2018quf} in the zero picture. 
Moreover, since we have  proven the classical equivalence of two models, the quantum equivalence will also hold, if we  perform  equivalent  integrable quantisations. We present the resulting quantum equivalence in section \ref{quantum}.

\section{The Strict Non-Compact Model}
\label{strictlimit}
The duality map provides us with a definition of a non-compact topological quantum field theory model,
before and after coupling to gravity. We also have a good understanding of a physical twisted
non-compact model, namely the twisted relevant deformations of $N=2$ Liouville theory,  as the limit of an integrable system \cite{Li:2018rcl,Li:2018quf}. In the present section, we establish the connection between these two systems in detail. We work at the level of the topological quantum field theories.

\subsection{The Non-Compact Correlators in the Zero Picture}
\label{scaling}

In the duality approach, we can compute all the correlators in the $-2$ picture in a 
non-compact topological quantum field theory model with a potential (\ref{noncompactsuperpotential})
with constant up to sub-subleading deformations of a leading monomial $Y^{-k}$. In principle, we have a solution for the generating function
of correlation functions $F_m^{(-2)}$. We would like to understand an equivalent description in a (more standard) zero picture.
To render the $0$ correlators well-defined, namely, to have a sensible time $t_0$ with a proper time evolution associated to it,
one adds to the superpotential (\ref{noncompactsuperpotential}) the linear term in $Y$. Clearly, this defines a new integrable system for which, among other quantities, the $0$ correlators as well as the $-2$ correlators make sense. We then define a limit on the free energy (in the spirit of \cite{Li:2018quf}, but slightly more general)  which 
leaves the $0$ correlators well-defined, yet eliminates the leading linear term in the potential. In this manner we define
$0$ correlators for the superpotential without the linear term. In the following, we study this limiting procedure.

\subsubsection{The Definition of the Zero Picture Correlators }

Firstly, we convince ourselves that the limit of the $0$-correlators is well-defined.
To that end, we need to understand the scaling of quantities with the parameter $\epsilon_1$ in which we take the limit.
The scaling reduces the dimension of the chiral ring (since the linear term in the superpotential will be eliminated) and is therefore clearly impactful.
We start with the superpotential $W_{lin}$ with linear term:
\begin{eqnarray}
W_{lin} &=& Y + v_0 + \frac{v_{-1}}{Y} + \dots + \frac{v_{-k}}{k Y^k} \, . \label{potentialwithlinearterm}
\end{eqnarray}
The scaling limit is defined as follows. We multiply $Y$ by $\epsilon_1$ and rescale all $u_{i \le -1}$ by $\epsilon_1$. By the formula for the $v_{-\alpha}$ in terms of the universal coordinates (see  \cite{Aoyama:1994nb,Li:2018quf}), this keeps all but the linear term in the potential fixed. 
In summary:
\begin{eqnarray}
Y & \longrightarrow & \epsilon_1 Y
\nonumber \\
u_{i \le -1} & \longrightarrow & \epsilon_1 u_{i \le -1}
\nonumber \\
v_0 & \longrightarrow & v_0 \, 
\nonumber \\
W_{lin} & \longrightarrow & W_{lin}(\epsilon_1) = \epsilon_1 Y + v_0 + \frac{v_{-1}}{Y} + \dots + \frac{1}{k Y^k} \, .
\end{eqnarray}
For $\epsilon_1 \neq 0$, a basis of the quotient ring is given by $1, Y^{-1}, \dots, Y^{-k}$.
For $\epsilon_1=0$, a basis of the quotient ring is given by $Y^{-2}, \dots, Y^{-k}$. The latter ring has dimension two less than the former.

The original model has a generating function of zero correlation functions, which we denote $F_m^{(0)}$. After the scaling
transformation, it depends on $\epsilon_1$. We would like to determine the behaviour of the generating function as a function
of the  parameter $\epsilon_1$ as we scale $\epsilon_1$ to zero.
We  think of $W_{lin}({\epsilon_1})$ as a constant
term plus a linear term in $\epsilon_1$. In this particular model, we have that $W_{lin}= L_{max}$ since the leading
power in the potential is one.  For the other formal power series,  $L_{min}$, the small $Y$ and
small $\epsilon_1$ expansions are straightforwardly compatible. We can work to linear order in $\epsilon_1$ at all stages.
We have for instance
\begin{equation}
W_{lin}(\epsilon_1) = \epsilon_1 Y + W \, ,
\end{equation}
where $W$ now indicates the target superpotential without linear term,
and
\begin{equation}
L_{min} = (k W_{lin})^{\frac{1}{k}}
\rightarrow 
  L (1 + \frac{\epsilon_1}{k} \frac{Y}{W} + 
 \dots) \, ,
\end{equation}
where $L$ is the formal series corresponding to the undeformed superpotential $W$. We can then use the formulas
for the zero picture one-point functions gathered in \cite{Aoyama:1994nb} for the rational model to understand the first derivatives of the scaled generator of correlation functions $F_m^{(0)} (\epsilon_1)$.
They are given by residues of fractional powers of the roots:
\begin{align}
\frac{\partial F_m^{(0)} (\epsilon_1)}{\partial u_0} =\langle \phi^0 \rangle_0 &= \frac{1}{2} \text{Res}_{Y=\infty} (L_{max}^2(\epsilon_1))\cr
&= \epsilon_1 u_0 u_{-1} + \frac{\epsilon_1^2}{2}\sum_{\alpha+\beta=k+2} u_{-\alpha}u_{-\beta} \,.
\cr
\frac{\partial F_m^{(0)} (\epsilon_1)}{\partial u_{-1}} =\langle \phi^{-1} \rangle_0 &= \text{Res}_{Y=\infty}\left( W \log \left(\frac{L_{max}(\epsilon_1)}{Y} - 1\right) \right) + \text{Res}_{Y=0} \left( \log \left(Y\, L_{min}(\epsilon_1) -\frac{1}{k} \right) \right)\cr
&= \frac{\epsilon_1}{2} u_0^2 + \epsilon_1^2\left( u_{-1} \log \epsilon_1 u_{-k}  +\ldots\right)
 \cr
\frac{\partial F_m^{(0)} (\epsilon_1)}{\partial u_{-\alpha}} =\langle  \phi^{-\alpha \le -2} \rangle_0 
&= \frac{1}{(\alpha+m-1)(\alpha-1)}
\text{Res}_{Y=0} (L_{min}(\epsilon_1))^{\alpha+k-1}\cr
&=\epsilon_1^2 \, \text{Res}_{Y=0} L^{\alpha+k-1} + O(\epsilon_1^3) \,. \label{limitonepointfunction}
\end{align}
We see that we have one linear term in $\epsilon_1$
in the generating function $F_m^{(0)}(\epsilon_1)$ of correlation functions, namely $\epsilon_1 u_{-1} u_{0}^2/2$, and otherwise quadratic
terms in $\epsilon_1$. We also deduce from the last equation
that the limit of the zero correlators for the operators that remain
in the spectrum, captured by the quadratic terms in $\epsilon_1$, is given by the naive formula, namely, in which we replace the root $L_{min}$ by the root $L$ of the
limit potential $W$.    We  conclude that, once we subtract the cubic term $\epsilon_1 u_{-1} u_{0}^2/2$, the limit
$\lim_{\epsilon_1 \rightarrow 0} F^{(0)}_m (\epsilon_1)$ has at most logarithmic divergences, and those are proportional to $u_{-1}$. The generating function of zero correlation functions that do not depend on $u_{-1}$
is given by integrating up the naive formula (\ref{limitonepointfunction}) of the one-point functions of the limit model.

\subsubsection*{A Scaling Law}

We can revisit the discussion more systematically by observing a scaling law for the model with linear term.
We recalled the first derivatives of the generating function $F_m^{(0)}$ of topological quantum field theory correlation functions (in the zero picture) in equations  (\ref{limitonepointfunction}). These first derivatives are determined algebraically. It is an interesting question whether the final integration of the generating function can also be performed algebraically. We already know this to be the case for the topological quantum field theories that arise
from deforming topologically twisted conformal field theories. Indeed, the latter satisfy the scaling equation
\cite{Dijkgraaf:1990qw,Li:2018quf}:
\begin{equation}
\sum_{j} (q_j-1) u_j \frac{\partial F_m^{(0)}}{\partial u_{-j}} = \left(\frac{c}{3}-3\right) F_m^{(0)} \, , \label{CFTscaling}
\end{equation}
where the sum is over all operators in the spectrum of the topological theory, $c$ is the central charge of the superconformal field theory and $q_i$ are the R-charges of the operator insertions. This equation  allows us to perform the final integration (by computing the left hand side to obtain the right hand side) for the generating function $F_m^{(0)}$ of correlation functions. Thus, for these theories, the calculation of the generating function can be performed completely algebraically. This equation is true both in the compact \cite{Dijkgraaf:1990qw} and the non-compact model \cite{Li:2018quf}. 

The question we turn to is whether there is a similar scaling law for the non-compact model with a linear term
(\ref{potentialwithlinearterm}) (or equivalently, a compact model with a $X^{-1}$ term). We propose the following answer: consider the generating function $F_m^{(0)}$. It has a single logarithmic term. Divide the argument of the logarithm, namely $u_{-k}$, by a scale factor $\mu$. Then, the generating function $F_m^{(0)}$  satisfies the anomalous scaling law:
\begin{equation}
\sum_{i=0}^k (q_i-1) u_{-i} \frac{\partial F_m^{(0)}}{\partial u_{-i}} =
-4 F_m^{(0)} + \sum_{i=1}^k \left(\frac{1}{2}+\frac{c}{6}\right) u_{-i}\frac{\partial F_m^{(0)}}{\partial u_{-i}}
+ \left(\frac{1}{2} + \frac{c}{6}\right) \mu \frac{\partial F_m^{(0)}}{\partial\mu} \, . \label{logarithmicscalinglaw}
\end{equation}
We have checked the anomalous scaling law in  examples at levels $k\le 7$. 

We now show that this scaling law is consistent with the limiting procedure towards the non-compact model without linear term, as well as the behaviour of the generating function in that limit. Indeed, in that limit, the terms quadratic in $u_{-1 \ge -i \ge -k}$ (that correspond to order $\epsilon_1^2$ terms in the generating function) will make for a right hand side in equation
(\ref{logarithmicscalinglaw}) equal to
\begin{equation}
\left(-4 + 2 \times \left(\frac{1}{2}+\frac{c}{6}\right)\right) F_m^{(0)} = \left(\frac{c}{3}-3\right) F_m^{(0)}
\end{equation}
matching onto the right hand side of equation (\ref{CFTscaling}). Thus, the limiting scaling law reproduces the known scaling law (\ref{CFTscaling}) for the deformation of the non-compact conformal field theory. 

The scaling law allows for an entirely algebraic determination of the generating function $F_m^{(0)}$ in the model
with linear term. It clarifies the identification of the quadratic terms as those corresponding to the zero time generating function of the non-compact model of interest.
We have illustrated 
these phenomena and the limiting procedure in low level examples  in appendix \ref{scalingandexamples}.

\subsubsection{The Relation Between the Zero and the Minus Two Pictures}
\label{picrelations}
In the previous subsections, we scaled out the linear term from the potential (\ref{potentialwithlinearterm}). The potential
then agrees  with the 
one from the integrable system obtained from duality. We have available a description of the correlators in both the $-2$ and
the $0$ picture and can now ask for the precise relation between these correlators. It is sufficient to observe that we 
know the relation between the universal coordinates, given by the equation
\begin{equation}
\tilde{u}^\beta=G^{\beta \alpha_0}(u) \, ,
\end{equation}
as well as the equality between the second derivatives of the generating function with respect to these coordinates \cite{Aoyama:1994nb}:
\begin{equation}
G^{\alpha \beta}= \partial_{u_\alpha} \partial_{u_\beta} F_m^{(0)}(u) = \partial_{\tilde{u}_\alpha} \partial_{\tilde{u}_\beta} F_m^{(-2)}(\tilde{u}) \, .
\end{equation}
Further derivatives give higher-point functions, and these derivatives can be related through the coordinate transformation
and the chain rule
\begin{equation}
d \tilde{u}_{\alpha} = \eta_{\alpha \gamma} \partial_{u_\beta} G^{-2,\gamma} d u_\beta
\qquad \partial_{\tilde{u}_\alpha} = (\eta_{\alpha \gamma} \partial_{u_\beta} G^{-2,\gamma})^{-1} \partial_{u_\beta} \, .
\end{equation}
For the three-point functions, we find for instance:
\begin{eqnarray}
\tilde{c}^{\alpha \beta \delta} (\tilde{u}) = (\eta_{\delta \gamma} \partial_{u_\epsilon} G^{-2,\gamma})^{-1} c^{\epsilon \alpha \beta}
= (\eta_{\delta \gamma} c^{\gamma \epsilon,-2})^{-1} c^{\epsilon \alpha \beta} (u) \, .
\end{eqnarray}
We can confirm this equation using  the relation between operators
\begin{equation}
\label{linearphirelation}
\phi^\alpha \equiv {(c^{-2})^\alpha}_\beta \tilde{\phi}^\beta
\end{equation}
valid modulo $W'$, as well as the  residue formulas for the correlators (see \cite{Aoyama:1994nb} for details).
 Thus, all correlators in the two pictures are in principle  related in a straightforward manner.

\subsection{The Link to the Strict Non-Compact Model}
\label{strictnoncompactsubsection}
Finally, we take the limit towards the strict non-compact models described in \cite{Li:2018rcl,Li:2018quf} with superpotential
\begin{equation}
W_{s.n.c} = \frac{v_{-k}}{k} Y^{-k} +\frac{v_{-k+1}}{k-1} Y^{-k+1}+ \dots + \frac{v_{-2}}{2} Y^{-2} \, .
\end{equation}
These are obtained by restricting to the topological degrees of freedom that arise from the physical Hilbert space of the twisted $N=2$ Liouville theory at radius $\sqrt{k \alpha'}$ \cite{Li:2018rcl,Li:2018quf}.
We need to eliminate
the constant term and the term proportional to $Y^{-1}$ in the superpotential (\ref{noncompactsuperpotential}).
We therefore send $v_0=u_{0}=\epsilon_2$ 
and $v_{-1}=u_{-1}=\epsilon_2$ to zero.  Given that the superpotential and the non-compact root $L$ behave regularly
in the limit, there is no subtlety in defining the limiting expressions. Those give rise to the models of \cite{Li:2018rcl,Li:2018quf}.
To recuperate all parameters present in \cite{Li:2018rcl,Li:2018quf}, one needs to restore two  parameters, which one can accomplish by rescaling $Y$ (to obtain a non-trivial parameter in front of the leading order term $Y^{-k}$) as well as shift the
variable $Y^{-1}$ (to find a non-trivial subleading term proportional to $Y^{-k+1}$). We refer to \cite{Li:2018rcl,Li:2018quf} for a full description of the resulting integrable system.

\section{Quantum Non-Compact Gravity}
\label{quantumsolution}
\label{replacement}
\label{quantum}

In this section, we comment on the solution of non-compact matter coupled to topological gravity on higher Riemann
surfaces.
Our strategy for solving the quantum model is simply to  exploit the solution to the quantum compact model through the equivalence
map.
Thus, 
we immediately describe the quantum theory with respect to the  non-compact reference time  $t_{-2}$. Schematically, the reasoning is that we turn the statements about the classical symplectic structure into equivalences on the quantum commutators, through the quantization:
\begin{eqnarray}
\{ X,t_0^c \} = 1 & \Longrightarrow & X = i \partial_{t_0^c} 
\nonumber \\
\{Y^{-1} , t_{-2} \} =1 & \Longrightarrow & Y^{-1} = i \partial_{t_{-2}}
\end{eqnarray}
in respectively the compact and the non-compact theory. The first equation leads to the standard quantised (or dispersionful) KdV hierarchy, while the second is its image under the equivalence map. We change variables on the left, as in the classical theory, and then  quantize and obtain the operators on the right (with identical operator ordering prescriptions in both theories). We refer to \cite{Dijkgraaf:1991qh} for a review of the quantum
KdV hierarchy in the context of topological gravity. There it is discussed that at zero compact times
$t_c^{i \ge 1}$, the initial condition for the quantum Lax operator $L_c$ appropriate for the topological quantum field theory coupled to gravity is
\begin{equation}
L_c(t^c_0) = \partial_{t^c_0}^{k_c} + t^c_0 \, ,
\end{equation}
because of the three-point function for primaries $X^i$ at the conformal point, fixed by charge conservation. This then uniquely determines
the $\tau$ function which is the generator of correlation functions for compact matter coupled to topological gravity:
\begin{equation}
\log \tau_c = \langle \exp \sum_{i \ge 0} t^c_i \phi_c^i \rangle \, .
\end{equation}
We record what these statements become under the equivalence map. 
The initial condition will map to
\begin{equation}
L(t_{-2}) = \partial_{t_{-2}}^k + t_{-2} \, .
\end{equation}
To understand  the initial condition, we need to realize that the relevant topological quantum field theory three-point functions are the correlators in the minus two picture. As we saw earlier, the operators
$\tilde{\phi}^i$ indeed have the same (minus two picture) three-point functions as do the $\phi_c^i$ operators
(in the zero picture).
Continuing in this vein, the time variables $t_i$ will couple to the $\tilde{\phi}^i$ operators
and we define the logarithm $\log \tau$ of the tau function as:
\begin{equation}
\log \tau = \langle \exp \sum_{ i \le -2} {t}_i \tilde{\phi}^i \rangle \, .
\end{equation}
Under the duality map, we then have the equality of tau-functions
\begin{equation}
\tau_c (t_i^c) = \tau (t_{-i-2}) \, .
\end{equation}
We can ask for the relation with the generator of correlation functions defined in terms of the original operators $\phi^i$ or
$\phi^\alpha$ in the non-compact model. We imagine we can restrict to the latter (by the decomposition of descendants into primaries and the renormalization of primary times).
We then observe that the fields $\phi^\alpha$ and $\tilde{\phi}^\alpha$ are related in the non-compact model by a linear transformation (see equation \eqref{linearphirelation}). We thus have a description of the correlators of the $\phi^\alpha$ (and $\phi^i$) fields in the zero picture as well. Thus, we have provided an algorithm for calculating the correlators of  quantum non-compact gravity.

\section{On Analytic Continuation}
\label{analyticcontinuation}
\label{analytic}
In this section, we discuss the extent to which our non-compact topological models are related to the compact topological models by analytic continuation. It is  interesting to perform analytic continuation in correlation functions computed for general positive compact level towards negative levels. Firstly, in \cite{Witten:1991mk}, this was shown to reproduce, at negative level $k_c=-1$, the Penner model for Euler characteristics of moduli spaces of Riemann surfaces. Secondly,
at level $k_c=-2$, a connection to unitary matrix models was uncovered \cite{Brezin:2010ar}.
Thirdly, at generic negative level $k_c$, an intriguing connection
to the spectral density of the $SL(2,\mathbb{R})/U(1)$ coset conformal field theory
was suggested in \cite{Brezin:2012uc}. 
Fourthly, one can make a tentative link to matrix models 
with negative power monomial potential \cite{Mironov:1994mv}. Here, we point out that analytic continuation reproduces a few elementary results in the non-compact topological theories that we obtained through duality.
However, we also show that generically, analytic continuation will lead to a different model. Thus, we situate our non-compact model more clearly with respect to the literature.

The section is structured as follows. We first make the  point that it is hard to generically relate our non-compact models to compact models through analytic continuation. Then, we backtrack and show that for a number of elementary results, there is a connection using analytic continuation. Finally, we illustrate how these links break down for generic correlators in a simple example.

\subsection{A Generic Argument}
\label{genericargument}
In the compact model coupled to topological gravity, it was argued in \cite{WittenAlgebraic} that the generator of 
correlation functions $F$ is a polynomial of maximal degree $k_c+1$ at genus zero. This follows from the charge conservation rule
\begin{equation}
s-3 = -\frac{c}{3} + \sum_{m=1}^s q_m
\end{equation}
where $s$ is the number of primary insertions and $q_i$ is their R-charge. The central charge $c$ equals $c=3-6/k_c$ where $k_c$ is a positive integer for compact minimal models. The order
of the polynomial is given by the maximal number of insertions. To maximize the number of insertions, we must maximize their R-charge (since the R-charge is strictly smaller than one). Thus, to compute the order of the polynomial, we solve the equation
\begin{equation}
s_{max} -3 = -\frac{c}{3}+ s_{max} \, q_{max} \,.
\end{equation}
For the maximal R-charge $q_{max}=1-2/k_c$ present in $N=2$ minimal models, we find that the maximal number of insertions is $s_{max}=k_c+1$, as stated. For non-compact models at central charge $c=3+6/k$, the same R-charge conservation rule holds, and one can reason similarly. In order to come as close to the compact model as possible, we will not allow the marginal deformation with R-charge $1$. If we allow the subleading R-charge $1-1/k$, then we will find a polynomial of order $2k-1$. If we only allow for the (sub-subleading) maximal R-charge $1-2/k$, then the polynomial will be of order $k-1$.\footnote{It is instructive to track how the duality evades this reasoning. In fact, a new vacuum $| \mbox{vac} \rangle_{(-2)} = \phi^{-2} | \mbox{vac} \rangle_{(0)}$ is defined in the $-2$ picture which changes R-charge such that in the charge conservation rule the non-compact central charge is transmuted  into the compact central charge. Moreover, the spectrum of R-charges of the tilded operators also matches the spectrum of compact R-charges. This leads to a polynomial order for the generating function identical to the order in the compact model.}  Thus, we already  strongly suspect that generic (zero picture) correlation functions cannot match (under analytic continuation). This generic argument is convincing, but we will confirm it through more detailed reasonings in the following. To increase intrigue, we first discuss a few correlators in which analytic continuation does provide a good guide to non-compact correlators.

\subsection{A Few Elementary Correlators}
It is known that the three- and four-point functions of topological gravity (at vanishing times) plus the associativity equation
determine the generating function $F$ of correlation functions uniquely \cite{WittenAlgebraic}. Thus, one strategy to see to what extent analytic continuation reproduces the non-compact correlation functions is by  starting out with the comparison of low-point functions at the conformal point (i.e. with monomial superpotential and all times equal to zero), and to build up from there.

\subsubsection{The Three-point Functions}
When the times vanish, the zero-, one- and two-point functions in topological gravity are zero.
The three-point functions of three primaries in the compact model are (see e.g. \cite{Dijkgraaf:1990dj})
\begin{equation}
\langle \phi_c^{i_1} \phi_c^{i_2} \phi_c^{i_3} \rangle = \delta_{\sum_{m=1}^3 q_{i_m},1- \frac{2}{k_c}} \, . \label{threecompact}
\end{equation}
The delta-function is dictated by charge conservation.  In the non-compact model, we have \cite{Li:2018rcl,Li:2018quf}
\begin{equation}
\langle \phi^{i_1} \phi^{i_2} \phi^{i_3} \rangle = \delta_{\sum_{m=1}^3 q_{i_m},1+ \frac{2}{k}} \, . \label{threenoncompact}
\end{equation}
The delta-function on the right  hand side can be obtained by analytic continuation $k_c \rightarrow -k$ from the compact correlator. Indeed, this is a direct consequence of the relation between the compact and non-compact central charges of the underlying $N=2$ superconformal field theories.
Note that these correlators match the zero picture matter correlators:
\begin{equation}
\langle X^{k_c-2} \rangle_0 = 1 
\end{equation}
and
\begin{equation}
\langle Y^{-k-2} \rangle_0 =1 \, ,
\end{equation}
which also translate into one another under analytic continuation in the level. This indicates that if analytic continuation is to work, it will be in the zero picture.

Thus, we seem to find that three-point functions compare well under analytic continuation. However, it is crucial to think about the spectrum of R-charges as well. In other words, one should also wonder about how to match  observables. The spectrum of R-charges in the zero picture  is the set $\{ 2/k,3/k,\dots,1 \}$ in the (strict) non-compact theory, and does not match the spectrum of R-charges $\{0,1/k,\dots,1-2/k \}$ in the compact theory. We will come back to this point, since it spoils the correspondence between the generating functions $F$ at low order despite the neat
continuation from equation (\ref{threecompact}) to equation (\ref{threenoncompact}).

\subsubsection{Four-point Functions}
A less trivial comparison is provided by the calculation of the four-point function on the sphere at zero times.
For the compact four-point function 
we have the charge conservation rule:
\begin{equation}
\sum_{m=1}^4 q_m = 2-\frac{2}{k_c} \, .
\end{equation}
The four-point function was calculated in the integrable system formalism in \cite{WittenAlgebraic}, by perturbing the superpotential to first order in times, and following the prescriptions  for computing the perturbed three-point functions to linear order in time. In particular, one uses the residue formula for the three-point function and the Hamiltonians and operators to linear order in time.
The result for the perturbed three-point function \cite{WittenAlgebraic}, linear in times, in our conventions reads  
\begin{eqnarray}
 c^{i_1 i_2 i_3}(t) &=&
 \delta_{i_1+i_2+i_3, k_c-2} 
 - \delta_{i_4+i_1+i_2+i_3,2 k_c-2} \, t_{i_4}^c  \times \nonumber \\
 & & \Big(i_4 - (i_1+i_4-k_c+1) \theta(i_1+i_4-k_c+1)  -(i_2+i_4-k_c+1) \theta(i_2+i_4-k_c+1)
 \nonumber \\
 & &
 -(i_3+i_4-k_c+1) \theta(i_3+i_4-k_c+1) \Big) + O(t^2) \, ,
 \end{eqnarray}
 where a sum over $i_4$ is implied.
At zeroth order in time, we confirm the three-point functions. The term linear in time fixes the
four-point function at zero time, which satisfies $\sum_{j=1}^4 i_j =2k_c-2$:
\begin{eqnarray}
\langle \phi^{i_1}_c \phi^{i_2}_c \phi^{i_3}_c \phi^{i_4}_c \rangle &=&  -i_4 +(i_1+i_4-k_c+1) \theta(i_1+i_4-k_c+1) \nonumber \\
&&
 + (i_2+i_4-k_c+1) \theta(i_2+i_4-k_c+1) \nonumber \\
&&
 + (i_3+i_4-k_c+1) \theta(i_3+i_4-k_c+1) 
 \nonumber \\
 &=& - \mbox{min} \,  \{ i_m, k_c-1-i_m \} \, .
\end{eqnarray}
The last equation is proven on a case by case basis.

We turn to the non-compact four-point function on the sphere at zero times.
At zero time, we take the conformal model
\begin{equation}
W(0) = \frac{1}{k} Y^{-k} \, .
\end{equation}
We  compute the intermediate results:
\begin{eqnarray}
W &=& \frac{Y^{-k}}{k} + \sum_{p=1}^{k-2} t_{-k+p} Y^{-k+p} + O(t^2)
\nonumber \\
\phi^{-i} &=& Y^{-i} +  \sum_{p=1}^{i-2} t_{-k+p} Y^{-i+p} (i-p-1)+ O(t^2) \, .
\end{eqnarray}
We use the residue formula 
\begin{equation}
c^{-l_1,- l_2,- l_3} =  \text{Res}_{Y} \left( \frac{\phi^{-l_1} \phi^{-l_2} \phi^{-l_3}}{W_Y} \right)
\end{equation}
for the three-point function as a function of times $t_i$, and wish to compute it to linear order in times in order to find the four-point function at zero times. We find:
 \begin{eqnarray}
c^{-l_1,- l_2,- l_3} &=& \delta_{l_1+l_2+l_3,k+2}
- \delta_{l_1+l_2+l_3+l_4,2k+2} \, t_{-l_4} \times
\nonumber \\
&& \Big( l_4-(l_1+l_4-k-1) \theta(l_1+l_4-k-1)
 -(l_2+l_4-k-1) \theta(l_2+l_4-k-1)
 \nonumber \\
 & & 
 -(l_3+l_4-k-1) \theta(l_3+l_4-k-1) \Big)
\end{eqnarray}
where we established the charge conservation equation for the four-point function
\begin{equation}
\sum_{m=1}^4 l_m = 2k+2   \, .
\end{equation}
In the end, we obtain a four-point function:
\begin{equation}
\langle \phi^{-l_1} \phi^{-l_2} \phi^{-l_3} \phi^{-l_4} \rangle = -\mbox{min} \, \{ l_m,k+1-l_m \} \, .
\end{equation}
We rewrite the correlators in the compact case as
\begin{equation}
\langle \phi^{i_1}_c \phi^{i_2}_c \phi^{i_3}_c \phi^{i_4}_c \rangle = 
 -k_c  \, \mbox{min} \, \{ q_{i_m},\frac{1}{2}+\frac{c_c}{6}-q_{i_m} \}
\end{equation}
which is proportional to a minimum of NS and R-sector R-charges. For the non-compact case, we write similarly:
\begin{equation}
\langle \phi^{i_1} \phi^{i_2} \phi^{i_3} \phi^{i_4} \rangle = 
 -k \, \mbox{min} \, \{ q_{i_m}, \frac{1}{2}+ \frac{c}{6}-q_{i_m} \}
\end{equation}
which has the same dependence on the charges, while the central charge $c$ is an analytic continuation of the compact central charge $c_c$. The overall sign can be made to agree by a change of sign convention for the deformation
times. The big caveat however is  that the spectrum
of charges does not match, as remarked earlier.

\subsection{An Explicit Difference}
In this subsection, we illustrate in  detail how the analytically continued compact and the non-compact topological gravity model part ways. We already mentioned the uniqueness of the higher point functions given
limited data on the lower point functions. This uniqueness theorem goes through mostly unchanged in the non-compact setting. Thus, to understand the difference between the analytically continued compact model and
the non-compact model it is indeed sufficient to study low-point functions. 
As we have already hinted at, the hiccup lies in the spectrum of R-charges (combined with anomalous R-charge conservation) which leads to differing low-point correlation functions. 

\subsubsection*{An Example }
We provide an example in which the reconstruction of the generating function differs for the compact and the non-compact case, showing non-uniqueness, even in the face of seeming analytic continuation. 
Consider the level $k_c=3$. For the compact case, we have the results
\begin{eqnarray}
W = \frac{X^3}{3} +u_1 X +u_0  \, &,& 
\qquad W_X = X^2 + u_1 \nonumber \\
\phi^0 = 1 \, &,&
\qquad
\phi^1 = X \, 
\end{eqnarray}
and 
\begin{equation}
F_m^{c} = \frac{u_1 u_0^2}{2} - \frac{u_1^4}{4!} \, .
\end{equation}
for the generating function of matter correlation functions in terms of the times $t_\alpha=u_\alpha$. Indeed, we have
a three-point function $\langle \phi^0 \phi^0 \phi^1 \rangle_{0,t=0}=\langle X \rangle_{0,t=0}=1$ at zero times, and we have the three-point function $\langle \phi^1 \phi^1 \phi^1 \rangle_{0,t} = \langle X^3 \rangle_{0,t} = - \langle u_1 X \rangle_{0,t} =-u_1$ at non-zero time, giving rise to the quartic term in the generating function. 

For the non-compact case, at level $k=3$, we have:
\begin{equation}
W = \frac{Y^{-3}}{3} + v_{-1}  Y^{-1} + v_0 \, ,
\qquad
W_Y = -Y^{-4} - v_{-1} Y^{-2}
\, .
\end{equation}
To find an analytic continuation map, we want to work in the zero picture, as argued previously. We  can, for instance, choose a basis  of  operators:
\begin{equation}
\phi^0= 1 \, ,
\qquad
 \phi^{-1} = Y^{-1} \, ,
\end{equation}
but these have no non-zero two- or three-point functions in the zero picture at zero times and cannot match the compact model. On the other hand, we might choose the
basis of operators:
\begin{eqnarray}
\phi^{-2} &=& Y^{-2} \, ,
\nonumber \\
\phi^{-3} &=& Y^{-3}
\end{eqnarray}
and we have a topological quantum field theory two-point function between these two operators in the
zero vacuum. (The two-point function  however is zero after coupling to topological gravity.) At zero times,
there is however no non-zero three-point function involving both types of operators, and therefore, again, we cannot match the compact  picture correlators. We conclude that the generating function  is not an analytic continuation of the compact generating function. This can also be verified using their explicit expressions.\footnote{See e.g. equation (\ref{Fsnc3}).} The underlying reason is that, in the zero picture, the spectrum of R-charges (as well as the anomalous R-charge contribution) in the compact model and the non-compact model differ. That makes (for instance) for different cubic terms in the generating function $F$. At higher levels, one finds even more manifest disagreement, for instance in the degree of the polynomial generating function, as argued in subsection \ref{genericargument}.

We conclude that the formal agreement of zero picture correlation functions that we obtained (at zero times) by analytic continuation, does not translate into identities for the generating functions. Of course, one can mend this disagreement between the compact and non-compact models, through duality. As we saw, in that case we identify the levels, without a sign flip, and do find a correspondence with the minus two picture of the non-compact theory, as described in detail in section \ref{duality}.

\subsubsection*{Summary}
Analytic continuation of compact correlators leads to  interesting results, including relations to known 
 integrable models \cite{Witten:1991mk,Brezin:2010ar,Brezin:2012uc,Mironov:1994mv}. Those models differ from the non-compact models we obtained by twisting $N=2$ Liouville theory at asymptotic radius $\sqrt{k \alpha'}$ \cite{Li:2018rcl,Li:2018quf}, and from the models we obtained  through duality.

\section{Conclusions}
\label{conclusions}

We have exploited the transformation of variables $X = Y^{-1}$ to solve
non-compact topological quantum field theories, before and after coupling them to topological gravity,
and on Riemann surfaces of arbitrary genus. The transformation reduces the problem to its compact
counterpart, which has been solved previously.  While the conceptual framework is simple, the details are slightly involved. We demonstrated that the duality maps compact zero picture correlators to non-compact minus two picture correlators. The minus two picture non-compact correlators are in turn related to their zero picture counterparts. Finally, the latter naturally arise from twisted topological conformal field theories as described in \cite{Li:2018rcl,Li:2018quf}. As a by-product, we were led to conjecture a scaling law for rational models with
a leading linear term.

Our duality has interesting conceptual consequences. Firstly, we shed new light on the solution of the topological quantum field theories discussed in \cite{Li:2018rcl}. Secondly, we extend the solution to non-compact gravity proposed in \cite{Li:2018quf} to arbitrary genus. Thirdly, 
the duality map provides  insight into the observation of \cite{Li:2018quf} that the critical exponents of non-compact two-dimensional gravity are the same as those of the compact models. Indeed,  the duality map implies that this must be the case. Non-compact matter in the presence of topological gravity seems to disturb the Riemann surface to a high degree, and precisely such that gravity compensates to make the matter degrees of freedom behave like compact matter once more.  

A further conceptual clarification of the gravitational backreaction of the non-compact matter  would be welcome. How does it precisely come about that the gravitational backreaction makes sure that the combined non-compact gravitational system has critical behaviour that matches the compact critical behaviour? Can this be reproduced by a lattice simulation? Further insight into this mechanism would clarify whether we should expect a similar phenomenon for matter of central charge $c>1$ coupled to ordinary gravity. That would solve a longstanding problem in two-dimensional gravity, i.e. it could foreshadow a stable endpoint for two-dimensional gravity coupled to more than minimal matter. Fourthly, the solution of the non-compact gravitational model is one key to solving topological string theories on asymptotically linear dilaton spaces which form a large class of analogues of non-compact Calabi-Yau manifolds. We look forward to exploiting the solution of the non-compact models further.

\section*{Acknowledgments}
It is a pleasure to thank our colleagues  for creating a stimulating research environment.

\appendix

\section{A Map and Illustrations}

This appendix is dedicated to  details and illustrations that  aid in improving our understanding of  aspects of the duality and the models we describe in the bulk of the paper.
\label{illustrations}
\subsection{The Map of the Universal Coordinates}
\label{universalcoordinatesmap}

Since we proved the classical equivalence of the compact and non-compact models, we can use the  formulas valid for the compact  topological quantum field theory \cite{Dijkgraaf:1990dj} in order to reconstruct the solution of the non-compact topological quantum field theory \cite{Li:2018rcl,Li:2018quf}. As argued in the bulk of the paper, the universal coordinates natural in the duality map differ from those typically used in the non-compact models.
Thus, it is useful to compute the coordinate change. 

The coordinate change can be constructed as follows. The  series expansion of the (compact model) variable $X$ in terms of $1/L_c$ at large $L_c$ is identical to the series expansion of $1/Y$ in terms of $1/L$ at large $L$, under the duality map. The subtlety lies in the fact that for the non-compact system the universal coordinates are defined in terms of the series expansion of $Y$ at large $L$. Still this information is sufficient to
find the link between the universal coordinates. For simplicity, we restrict to the case where
$u^{-2}=u_{-k}=1$ and $u^{-3}=u_{-k+1}=0$.\footnote{These parameters can be restored by respectively rescaling
and shifting the variable $Y^{-1}$.} We then have the non-compact universal coordinates
defined by the expansion
\begin{equation}
Y = \frac{1}{L} \left(1+ \sum_{i=2}^\infty u^{-i-2} L^{-i}\right)
\end{equation}
and therefore derive 
\begin{align}
    \frac{1}{Y} &= L \left(1+ \sum_{i=2}^\infty u^{-i-2} L^{-i}\right)^{-1} \cr
    &= L \left(1 + \sum_{k=1}^\infty (-1)^k \left(\sum_{i=2}^\infty u^{-i-2} L^{-i}\right)^k \right)\, .
\end{align}
Comparing the latter sum to the compact universal coordinates 
\begin{equation}
X = L_c \left(1 - \sum_{i=2}^\infty u^{i-2}_c L_c^{-i}\right)
\end{equation}
by replacing $Y^{-1}=X$ and $L \leftrightarrow L_c$, we find the relation between the universal coordinates in the two systems:
\begin{equation}
-u_c^{a-2} = \sum_{k=1}^\infty (-1)^k \sum_{\sum_{j=1}^k a_j= a, a_j \ge 2}  \prod_j  u^{-a_j-2} \, .
\end{equation}
We can also compute the inverse relation, expressing the non-compact universal coordinates in terms of the compact universal coordinates:
\begin{equation}
u^{-a-2} = \sum_{k=1}^\infty \sum_{\sum_{j=1}^k a_j= a, a_j \ge 2}  \prod_j  u^{a_j-2}_c \, .
\end{equation}
Thus, we have  established an explicit map between the compact and the non-compact models for
the standard universal coordinates. Note that this also establishes a map between
tilded  universal coordinates in the non-compact model used in the bulk of the paper and the untilded
universal coordinates for the non-compact model used in \cite{Li:2018rcl,Li:2018quf}.


\subsection{The Equivalence Exemplified}
\label{example}
We compare  topological quantum field theory generating functions of correlations functions. On the one hand, we recall the
compact generating function $F^{(0)}_{c,m}$ in terms of the compact universal coordinates $u_c$, and on the other hand calculate the non-compact
topological quantum field theory generating function $F_m^{(-2)}$ of $\alpha_0=-2$ correlators,  in terms of  the $\alpha_0=-2$
universal coordinates $\tilde{u}$. We show that the latter function $F_m^{(-2)}$ coincides with the function $F^{(0)}_{c,m}$ after a relabelling. 
We also illustrate the coordinate map of appendix \ref{universalcoordinatesmap} concretely at levels three and four.

At level $k_c=3=k$, we have the potentials:
\begin{eqnarray}
W_c &=& \frac{X^3}{3} + v_1 X + v_0
\nonumber \\
W &=& \frac{Y^{-3}}{3} + v_{-1} Y^{-1} + v_0 \, ,
\end{eqnarray}
and after some calculation, we find the generators of correlation functions:
\begin{eqnarray}
F^{(0)}_{m,c} &=& \frac{1}{2} (u_0^c)^2 u_1^c-\frac{1}{24} (u_1^c)^4
\nonumber \\
F_m^{(-2)}(\tilde{u}^\alpha) &=& \frac{1}{2} (\tilde{u}_{-2})^2 \tilde{u}_{-3} - \frac{1}{24} (\tilde{u}_{-3})^4  \, . \label{gen3}
\end{eqnarray}
We have used the intermediate result that
\begin{eqnarray}
G^{-2,-2} &=& u_{-1}
\nonumber \\
G^{-2,-3} &=& u_0
\end{eqnarray}
which codes the  relation between the non-compact universal coordinates. Moreover, the duality relates the tilded universal non-compact
coordinates $\tilde{u}$ to the compact universal coordinates $u^c$. We have therefore
\begin{eqnarray}
 u_1^c \equiv \tilde{u}_{-3} = u_{-1} && 
\nonumber \\
u_0^c \equiv \tilde{u}_{-2} = u_0  \, , &&
\end{eqnarray}
and the generating functions (\ref{gen3})  coincide, as implied by duality.

Next, 
we put $k_c=4=k$, and work with the superpotentials:
\begin{eqnarray}
W_c &=& \frac{X^4}{4} + v_2 X^2 + v_1 X+v_0
\nonumber \\
W &=& \frac{Y^{-4}}{4} + v_{-2} Y^{-2} + v_{-1} Y^{-1} + v_0 \, .
\end{eqnarray}
After computing the formal series $L_c$ and $L$, and the operators $\phi^\alpha$, and plugging them into the second
derivative potentials $G^{\alpha \beta}$, we can integrate up twice to find the compact and non-compact generating functions. Alternatively, we can use the known one-point functions and a scaling relation.
A number of lines of calculation later one finds:
\begin{eqnarray}
F^{(0)}_{m,c}(u_c) &=& \frac{u_0^c (u_1^c)^2}{2} + \frac{(u_0^c)^2 (u_2^c)}{2} - \frac{(u_1^c)^2 (u_2^c)^2}{4}+\frac{(u_2^c)^5}{60} 
\nonumber \\
F_m^{(-2)}(\tilde{u}) &=& \frac{\tilde{u}_{-2} \tilde{u}_{-3}^2}{2}
+ \frac{\tilde{u}_{-2}^2 \tilde{u}_{-4}}{2}
-\frac{\tilde{u}_{-3}^2 \tilde{u}_{-4}^2}{4}
+ \frac{\tilde{u}_{-4}^5}{60} \, . \label{gen4}
\end{eqnarray}
The universal coordinates map as
\begin{eqnarray}
&& u^c_0 \equiv \tilde{u}_{-2} = u_{0}- \frac{u_{-2}^2}{2} = G^{-2,-4}
\nonumber \\
&& u^c_1 \equiv \tilde{u}_{-3} = u_{-1}
\nonumber \\
&& u^c_2 \equiv \tilde{u}_{-4} =  u_{-2} \, .
\end{eqnarray}
In appendix \ref{scalingandexamples} we will independently recover these results from the generating function of the non-compact model calculated in the zero basis and using the limiting procedure detailed in the text. We end this section by observing that the expression for the generating functions $F^{(0)}_{m,c}(u_c)$ and $F_m^{(0)}(\tilde{u})$ in equation (\ref{gen4}) agree,
as implied by  duality.

\section{Models and  Scaling Laws}
\label{scalingandexamples}
\label{modelsandscaling}
In this appendix, we provide  examples of the limiting procedure discussed in subsection \ref{scaling} of the paper as well as illustrations of the  scaling law for a non-compact model supplemented with a linear term in the potential.
In particular, we compute examples of the generating function $F_m^{(0)}$ through integration, and checked that the result agrees with the algebraic calculation of the function $F_m^{(0)}$  using the proposed scaling law (\ref{logarithmicscalinglaw}). We also explicitly provide the limiting form of the generating functions
in the 
limit $\epsilon_1 \rightarrow 0$, defined in subsection \ref{scaling} in the bulk of the paper, as well as an example calculation of the relation between different sets of universal coordinates.
 
 As described after equation (\ref{limitonepointfunction}), in the generating function $F_m^{(0)}$ of the model
 with the linear term, we scale all the universal coordinates $u_{-j}$ by $\epsilon_1$ except $u_0$ and then the generating function of the strict non-compact theory is the one obtained by extracting the $O(\epsilon_1^2)$ coefficient, along with setting $u_0 = u_{-1} = \epsilon_2=0$. One can then check that the resulting generating function indeed satisfies the conformal field theory scaling \eqref{CFTscaling} with non-compact central charge $c=3+\frac{6}{k}$. Below we list the zero picture generating function, its $\epsilon_1$-expansion and finally, the generating function of the strict non-compact model\footnote{For $k=1,2$, the generating function for the rational model in the zero picture has been obtained in \cite{Aoyama:1994nb}. } in the zero picture for levels $k=3,4$ and $5$.

\subsubsection*{Level Three}
\begin{align}
\label{Fk3}
F_m^{(0)} &= \frac{1}{2} u_{-1}^2 \log \left(\frac{u_{-3}}{\mu }\right)-\frac{u_{-2}^4}{12 u_{-3}^2}+\frac{u_{-1} u_{-2}^2}{2 u_{-3}}+u_{-3} u_0 u_{-2}+\frac{u_{-3}^3}{6}+\frac{1}{2} u_{-1} u_0^2  \, ,
\cr
F_m^{(0)} (\epsilon_1) &=\frac{ \epsilon_1 }{2} u_{-1} u_{0}^2+ \epsilon_1 ^2 \left(\frac{1}{2} u_{-1}^2 \log \left(\frac{\epsilon_1  u_{-3}}{\mu }\right)-\frac{u_{-2}^4}{12 u_{-3}^2}+\frac{u_{-1} u_{-2}^2}{2 u_{-3}}+u_{-3} u_{0} u_{-2}\right)+ O(\epsilon_1^3) \, . \cr 
\end{align}
The generating function $F_{s.n.c.}$ of the strict non-compact matter model of subsection
\ref{strictnoncompactsubsection}
is obtained by setting $u_{-1}=u_0 = 0$ and taking the $\epsilon_1^2$ coefficient of $F_m$. This leads to the generating function $F_{s.n.c.}$ for the strict non-compact model
\begin{equation}
F_{s.n.c.} = - \frac{u_{-2}^4}{12u_{-3}^2}     \, .
\label{Fsnc3}
\end{equation}

\subsubsection*{Level Four}
\begin{align}
F_m^{(0)} =&-\frac{u_{-3}^6}{24 u_{-4}^4}+\frac{u_{-2} u_{-3}^4}{4 u_{-4}^3}-\frac{u_{-2}^2 u_{-3}^2}{2 u_{-4}^2}+\frac{1}{2} u_{-4}^2 u_{-3}+\frac{u_{-2}^3}{6 u_{-4}}+\frac{1}{2} u_{-1} u_0^2\cr
&+\left(\frac{u_{-3} u_{-2}}{u_{-4}}-\frac{u_{-3}^3}{6 u_{-4}^2}\right) u_{-1}+\left(\frac{u_{-3}^2}{2}+u_{-4} u_{-2}\right) u_0+\frac{1}{2} u_{-1}^2 \log \left(u_{-4}\right)
\cr
F_m^{(0)} (\epsilon_1) &= \frac{\epsilon_1}{2}   u_{-1} u_{0}^2+\epsilon_1 ^2\left(-\frac{u_{-3}^6}{24 u_{-4}^4}+\frac{u_{-3}^4 u_{-2}}{4 u_{-4}^3}-\frac{u_{-3}^2 u_{-2}^2}{2 u_{-4}^2}+\frac{u_{-2}^3}{6 u_{-4}}-\frac{u_{-1} u_{-3}^3}{6 u_{-4}^2}\right. \cr
&\left.+\frac{1}{2} u_{-1}^2 \log \left(\frac{\epsilon_1  u_{-4}}{\mu }\right)+\frac{1}{2} u_{0} u_{-3}^2+\frac{u_{-2} u_{-1} u_{-3}}{u_{-4}}+u_{-4} u_{-2} u_{0}\right) + O(\epsilon_1^3) \, .
\end{align}
The generating function of the strict non-compact model at level $k=4$ is 
\begin{equation}
F_{s.n.c.}=    -\frac{u_{-3}^6}{24 u_{-4}^4}+\frac{u_{-2} u_{-3}^4}{4 u_{-4}^3}-\frac{u_{-2}^2 u_{-3}^2}{2 u_{-4}^2}+\frac{u_{-2}^3}{6 u_{-4}} \, .
\end{equation}

\subsubsection*{Level Five}
\begin{align}
F_m^{(0)} =& -\frac{u_{-4}^8}{24 u_{-5}^6}+\frac{3 u_{-4}^6 u_{-3}}{10 u_{-5}^5}+\frac{u_{-4}^2 \left(8 u_{-3}^3+12 u_{-4} u_{-2} u_{-3}+u_{-4}^2 u_{-1}\right)}{12 u_{-5}^3}-\frac{u_{-4}^4 \left(15 u_{-3}^2+4 u_{-4} u_{-2}\right)}{20 u_{-5}^4}\cr
&+\frac{-u_{-3}^4-8 u_{-4} u_{-2} u_{-3}^2-4 u_{-4}^2 u_{-1} u_{-3}-4 u_{-4}^2 u_{-2}^2}{8 u_{-5}^2}+\frac{u_{-1} u_{-3}^2+u_{-2}^2 u_{-3}+2 u_{-4} u_{-2} u_{-1}}{2 u_{-5}}\cr
&+\frac{1}{2} \left(u_{-1}^2 \log \left(\frac{u_{-5}}{\mu }\right)+u_0^2 u_{-1}+2 u_{-4} u_{-3} u_0\right)+u_{-5} \left(\frac{u_{-4}^2}{2}+u_{-2} u_0\right) \, ,
\cr
F_m^{(0)} (\epsilon_1) =&\frac{\epsilon_1}{2} u_{-1} u_0^2 +\epsilon_1 ^2 \left(\frac{1}{2} u_{-1}^2 \log \left(\frac{u_{-5} \epsilon_1 }{\mu }\right)-\frac{u_{-4}^8}{24 u_{-5}^6}+\frac{3 u_{-3} u_{-4}^6}{10 u_{-5}^5}-\frac{\left(15 u_{-3}^2+4 u_{-4} u_{-2}\right) u_{-4}^4}{20 u_{-5}^4}\right.\cr
&\left.+\frac{\left(8 u_{-3}^3+12 u_{-4} u_{-2} u_{-3}+u_{-4}^2 u_{-1}\right) u_{-4}^2}{12 u_{-5}^3}
+\frac{-u_{-3}^4-8 u_{-4} u_{-2} u_{-3}^2-4 u_{-4}^2 u_{-1} u_{-3}-4 u_{-4}^2 u_{-2}^2}{8 u_{-5}^2}\right.\cr
&\left.+\frac{u_{-1} u_{-3}^2+u_{-2}^2 u_{-3}+2 u_{-4} u_{-2} u_{-1}}{2 u_{-5}}+u_{-5} u_{-2} u_0+u_{-3} u_0 u_{-4}\right)+O(\epsilon_1^3) \, .
\end{align}
The generating function of the strict non-compact model is given by
\begin{align}
F_{s.n.c.} =&    -\frac{u_{-4}^8}{24 u_{-5}^6}+\frac{3 u_{-3} u_{-4}^6}{10 u_{-5}^5}-\frac{\left(15 u_{-3}^2+4 u_{-4} u_{-2}\right) u_{-4}^4}{20 u_{-5}^4}+\frac{\left(8 u_{-3}^3+12 u_{-4} u_{-2} u_{-3}\right) u_{-4}^2}{12 u_{-5}^3}\cr
&\hspace{3cm}+\frac{u_{-3} u_{-2}^2}{2 u_{-5}}+\frac{-u_{-3}^4-8 u_{-4} u_{-2} u_{-3}^2-4 u_{-4}^2 u_{-2}^2}{8 u_{-5}^2} \, .
\end{align}
The generating functions of the strict non-compact models can be seen to agree with the results obtained in \cite{Li:2018rcl} and they satisfy the scaling law (\ref{CFTscaling}).

\subsubsection*{Relations between universal coordinates from the generating function}
\label{uutilde}

Once we have the generating function of the non-compact theory, it is a simple matter to take derivatives and obtain the two point function $G^{\beta \alpha_0}$. According to \cite{Aoyama:1994nb} this provides the relation between the universal coordinates $u$ and $\tilde{u}$ in the zero and minus two bases. The superpotential for which we do this is 
\begin{equation}
    W = \frac{Y^{-k}}{k} + v_{-k+2} Y^{-k+2}+\ldots + v_{-1} Y^{-1} + v_0 \,.
    \label{noncompactY2}
    \end{equation}
The generating function that has been calculated for the examples in section \ref{example} is for the rational model
\begin{equation}
    W = Y + v_0 + \ldots +\frac{v_{-k}}{k} {Y^{-k}}\,.
\end{equation}
As discussed in detail previously,  to obtain results in the non-compact model defined by the potential \eqref{noncompactY2} we perform the $Y\rightarrow \epsilon_1 Y$ scaling, accompanied by the appropriate scaling of the $u$-variables. In addition, in order to apply these results to the model defined by equation \eqref{noncompactY2}, we need to set $u_{-k}=1$ and $u_{-k+1}=0$. 

The relevant two point functions of the rational model that define the universal coordinates $\tilde{u}^{-\alpha}$ in the $-2$ picture, are given by
\begin{equation}
\tilde{u}^{-\alpha} = G^{-\alpha,-2}\big|_{u_{-k}=1, u_{-k+1}=0} = \frac{\partial^2 F_m^{(0)}}{\partial u_{-\alpha} \partial u_{-2}} \bigg|_{u_{-k}=1, u_{-k+1}=0}    \quad\text{for}\quad \alpha \in \{-2, -3, \ldots, -k\}
\end{equation}
Let us illustrate this in the case of the level three  model. The generating function is given by (see equation \eqref{Fk3}):
\begin{equation}
 F_m^{(0)} = \frac{1}{2} u_{-1}^2 \log \left(\frac{u_{-3}}{\mu }\right)-\frac{u_{-2}^4}{12 u_{-3}^2}+\frac{u_{-1} u_{-2}^2}{2 u_{-3}}+u_{-3} u_0 u_{-2}+\frac{u_{-3}^3}{6}+\frac{1}{2} u_{-1} u_0^2   \, .
\end{equation}
By differentiating with respect to the coordinates $u$ we find that
\begin{align}
   G^{-2,-2} &= \frac{u_{-1}}{u_{-3}}-\frac{u_{-2}^2}{u_{-3}^2} \,,  \cr
    G^{-2,-3} &= \frac{2 u_{-2}^3}{3 u_{-3}^3}-\frac{u_{-1} u_{-2}}{u_{-3}^2}+u_{0} \,.
\end{align}
Applying the limiting procedure discussed above, we obtain
\begin{equation}
    \tilde{u}^{-2} = u_{-1} \qquad \tilde{u}^{-3}= u_0 \,,
\end{equation}
confirming what we found using the Lax operators in appendix \ref{example}. 

The same procedure can be applied to the higher level examples.
For level four we find
\begin{equation}
    \tilde{u}^{-2} = u_{-2}\qquad \tilde{u}^{-3} = u_{-1}\qquad \tilde{u}^{-4} = u_0 - \frac{1}{2}u_{-2}^2\,.
\end{equation}
and for level five:
\begin{equation}
    \tilde{u}^{-2}= u_{-3}\qquad \tilde{u}^{-3}=u_{-2}\qquad \tilde{u}^{-4} = u_{-1}-u_{-3}^2 \qquad \tilde{u}^{-5} = u_0 - u_{-2}u_{-3} \,.
\end{equation}
Once the map between the universal coordinates in the minus two and zero pictures is obtained, one can proceed to relate all $n$-point correlators as discussed in subsection \ref{picrelations}.

\bibliographystyle{JHEP}

\end{document}